\documentclass[usenatbib,usegraphicx]{mn2e}

\usepackage{times}
\usepackage[dvipdfm]{hyperref}
\usepackage{amssymb}
\usepackage{bm}

\newcommand{\be}{\begin{equation}}
\newcommand{\ee}{\end{equation}}
\newcommand{\bea}{\begin{eqnarray}}
\newcommand{\eea}{\end{eqnarray}}

\title[Polarization from reconfinement shocks]
{Polarization of synchrotron emission from relativistic reconfinement shocks}

\author[K. Nalewajko]
{Krzysztof Nalewajko\thanks{E-mail: knalew@camk.edu.pl} \\
Nicolaus Copernicus Astronomical Center, Bartycka 18, 00-716 Warsaw, Poland}

\begin{document}

\maketitle

\begin{abstract}
We study the polarization properties of relativistic reconfinement shocks with chaotic magnetic fields. Using our hydrodynamical model of their structure, we calculate synthetic polarization maps, longitudinal polarization profiles and discuss the spatially averaged polarization degree as a function of jet half-opening angle $\Theta_j$, jet Lorentz factor $\Gamma_j$ and observer inclination angle to the jet axis $\theta_{obs}$. We find, that for $\theta_{obs}\lesssim\Theta_j$ the wave electric vectors are parallel in the vicinity of the structure ends and perpendicular in between, while for $\theta_{obs}>\Theta_j$ the polarization can only be perpendicular. The spatially averaged polarization degree does not exceed $30\%$. Parallel average polarization, with polarization degrees lower than $10\%$, have been found for $\theta_{obs}<\Theta_j$ under the condition $\Gamma_j\Theta_j>1$. As earlier works predicted the parallel polarization from relativistic conical shocks, we explain our results by discussing conical shocks with divergent upstream flow.
\end{abstract}


\section{Introduction}
\label{sec:intro}

A great variety of polarization properties has been found in relativistic AGN jets on different wavelengths (radio, mm and optical) and scales (subparsec through kiloparsec). From the theoretical point of view, high polarization degrees in this wavelenght range implicate the synchrotron emission mechanism taking place in the presence of ordered magnetic fields. The polarization electric vectors are being observed parallel or perpendicular to the jet axis. Perpendicular polarization, prevailent in the large-scale jets in radio \citep{1994AJ....108..766B} and optical \citep{2006ApJ...651..735P} bands, has been particularily difficult to explain, as it requires the dominance of longitudinal magnetic fields in the emission regions. In expanding jet the parallel (poloidal) component of the magnetic field decays faster than the perpendicular (toroidal) one, so one needs a process, in which poloidal component is amplified on large scales \citep{1984RvMP...56..255B}. The velocity shear at the jet boundary is usually invoked \citep{1981ApJ...248...87L}.

Chaotic (tangled) magnetic fields are thought to dominate the jets at distances larger than a few parsecs, but they must be statistically anisotropic to produce a net linear polarization \citep{1962SvA.....5..678K}. It has been pointed out by \cite{1980MNRAS.193..439L}, that initially isotropic chaotic magnetic field becomes anisotropic after crossing the shock front due to compression of plasma. Polarization degree of emission from compressed magnetic field has been calculated by \cite{1985ApJ...298..301H} and more general formulae, incorporating distortions due to velocity shear, have been given by \cite{1990MNRAS.242..616M} and \cite{2002MNRAS.329..417L}. Polarization from stationary relativistic conical shocks has been studied by \cite{1990ApJ...350..536C}, hereafter \citeauthor*{1990ApJ...350..536C}. They found that high degrees of parallel polarization may be obtained, but the degree of perpendicular polarization is limited to $\sim10\%$. Since the knots of blazar jets are observed with a perpendicular polarization of higher degree (e.g. \citealt*{2002ApJ...577...85M}), \cite{2006MNRAS.367..851C} introduced large-scale poloidal magnetic field component in the upstream flow.

In weakly magnetized AGN jets, a sequence of so-called reconfinement shocks forms, resulting from the interaction between the jet and the external medium \citep{1983ApJ...266...73S}. In our previous paper (\citealt{2009MNRAS.392.1205N}, hereafter \citeauthor*{2009MNRAS.392.1205N}) we studied the structure and energy dissipation efficiency of axisymmetric reconfinement shocks and compared our results to the analytical formulae given by \cite{1997MNRAS.288..833K}. We will now use a model developed there as a basis for calculating the polarization of emission originating behind the shock front from chaotic magnetic fields compressed at the shock. A reconfinement shock is treated as a set of conical shocks with inclination angle dependent on the position along the symmetry axis. The upstream flow is assumed to be expanding freely, thus it diverges. We begin with demonstrating the difference between conical shocks with parallel and divergent upstream flow.

This paper is organized as follows. In \S\ref{sec:oblique} we present the scheme for calculating the degree and positional angle of linearly polarized emission from oblique shocks. In \S\ref{sec:conical} we study the polarization from conical shocks with diverging upstream flow. In \S\ref{sec:reconf} we study the polarization from reconfinement shocks: the polarization maps are presented in \S\ref{subsec:reconf:resolv}, the longitudinal polarization profiles in \S\ref{subsec:reconf:semiresolv} and the spatially averaged polarization degrees in \S\ref{subsec:reconf:unresolv}. Our results are discussed and summarized in \S\ref{sec:discussion}.

We use a term 'perpendicular/parallel polarization', meaning the orientation of the wave electric vectors with respect to the jet axis. The primed quantities are those measured in the frame comoving with the downstream plasma, in contrast to the quantities measured in the external frame.

\section{Calculating the polarization from the oblique shock}
\label{sec:oblique}

In weakly magnetized shocks the cold upstream matter can dissipate a significant fraction of kinetic energy, which is partially transferred to a population of nonthermal relativistic electrons/positrons. Those particles emit synchrotron radiation and the most energetic of them are expected to cool rapidly enough, that the emission source is tightly localized in the shock vicinity. For a stationary relativistic shock, although the emitting elements can be moving with large Lorentz factor $\Gamma=(1-\beta^2)^{-1/2}$, $\beta=v/c$, the source position does not change in time. The relativistic enhancement of the intrinsically isotropic radiation is of the factor $\mathcal{D}^3/\Gamma$ \citep{1997ApJ...484..108S}, where
\be
\mathcal{D} = \frac{1}{\Gamma\left(1-\beta\,\cos\xi_{ke}\right)}
\ee
is the Doppler factor and $\xi_{ke}$ is the angle between the direction of the element's motion and the observer direction in the external frame. Considering only the bolometric luminosity, we neglect the spectral index here. Thus, the Stokes parameters in the observer frame are:
\bea
\nu I_\nu &=& \frac{\mathcal{D}^3}{\Gamma}(\nu I_\nu)'\,,\\
\nu Q_\nu &=& \Pi\,\cos(2\chi_E)\,\nu I_\nu\,,\\
\nu U_\nu &=& \Pi\,\sin(2\chi_E)\,\nu I_\nu\,;
\eea
where $\Pi$ is the degree of linear polarization and $\chi_E$ is the electric vector positional angle (EVPA). Like in most studies of polarization from astrophysical jets, we measure $\chi_E$ from the projected direction of the jet axis.

\begin{figure}
\includegraphics[width=\columnwidth,bb=200 95 325 270]{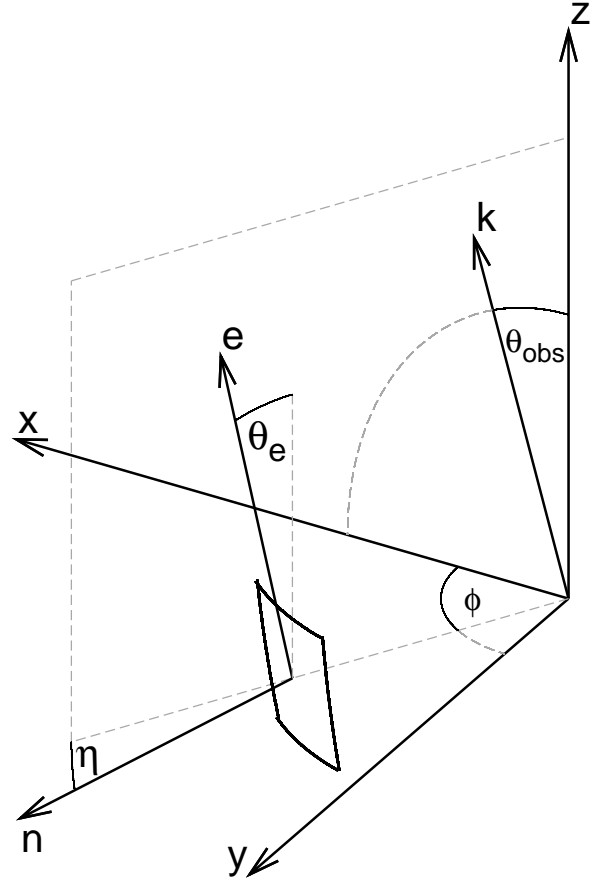}
\caption{The coordinate system used for calculating the polarization degree and the electric vector positional angle (EVPA). The jet direction is along the $z$-axis. The observer is located in the $xz$-plane, inclined to the jet direction by $\theta_{obs}$. \emph{The black contour} represents a shock surface element located at the positional angle $\phi$. Vector $\bm{n}$, normal to the shock element, is inclined to the local radial coordinate by $\eta$. The downstream velocity field direction $\bm{e}$ makes an angle $\theta_e$ with the jet axis. Since we assume axisymmetric jet, vectors $\bm{n}$ and $\bm{e}$ are aligned with the $\phi=\rm const$ plane.}
\label{fig2.1}
\end{figure}

In the following, we present our method of calculating $\Pi$ and $\chi_E$ for emission from oblique shocks with chaotic magnetic fields. We choose a cartesian coordinate system $(x,y,z)$, in which the jet axis is aligned with the $z$-axis and the observer is pointed by a unit vector $\bm{k}$ contained in the $xz$-plane. The inclination of the observer direction to the jet axis is $\theta_{obs}$. The shock front is inclined to the jet axis at angle $\eta$ and its normal vector $\bm{n}$ makes an azimuthal angle $\phi$ with the $xz$-plane. A downstream fluid element is characterized with velocity $\beta=v/c$ and the inclination of its velocity direction $\bm{e}$ to the jet axis $\theta_e$. Thus, we have (see Fig. \ref{fig2.1})
\bea
\bm{k} &=& [\sin\theta_{obs},0,\cos\theta_{obs}], \\
\bm{e} &=& [\sin\theta_e\cos\phi,\sin\theta_e\sin\phi,\cos\theta_e], \\
\bm{n} &=& [\cos\eta\cos\phi,\cos\eta\sin\phi,-\sin\eta], \\
\cos\xi_{ke} &=& \bm{k}\cdot\bm{e}=\sin\theta_{obs}\sin\theta_e\cos\phi+\cos\theta_{obs}\cos\theta_e.
\eea

We perform a Lorentz transformation into the fluid rest-frame $\mathcal{O}'$. The observer position vector transforms like
\be
\label{veckp}
\bm{k}' = \mathcal{D}\left\{\bm{k}+\left[(\Gamma-1)\,\bm{k}\cdot\bm{e}-\Gamma\beta\right]\bm{e}\right\}.
\ee
The inclination of the shock surface transforms like
\be
\tan(\eta'-\theta_e)=\Gamma\tan(\eta-\theta_e).
\ee
Using $\eta'$, we find the normal vector to the shock surface:
\be
\bm{n}' = [\cos\eta'\cos\phi,\cos\eta'\sin\phi,-\sin\eta'].
\ee
Now we adopt a formula from \citet{1985ApJ...298..301H} for the degree of polarization:
\be
\Pi=\frac{\alpha+1}{\alpha+5/3}\times\frac{(1-\kappa^2)\left[1-\left(\bm{k}'\cdot\bm{n}'\right)^2\right]}{2-(1-\kappa^2)\left[1-\left(\bm{k}'\cdot\bm{n}'\right)^2\right]},
\ee
where $1/\kappa$ is the shock compression ratio and $\alpha$ is the spectral index of optically thin synchrotron radiation: $F_\nu\propto\nu^{-\alpha}$. For the rest of this paper we will use $\alpha=0.5$, for which the maximum value of the polarization degree is $\Pi_{max}\sim0.7$.

The magnetic vector of the polarized electromagnetic wave in $\mathcal{O'}$ is both normal to the propagation direction (observer) and tangent to the shock surface, thus
\be
\bm{B}'\propto\left(\bm{k}'\times\bm{n}'\right).
\ee
We show in Appendix \ref{app1}, that the polarization angle is invariant in the Lorentz transformation, when measured with respect to the transformation vector, in this case $\bm{e}$. First, we introduce an orthogonal basis ($\bm{v}'$, $\bm{w}'$) in the plane of the sky, for example
\bea
\bm{v}' &=& \left[k_z',0,-k_x'\right],\\
\bm{w}' &=& \left[-k_x'k_y',1-k_y'^2,-k_y'k_z'\right].
\eea
Then, the positional angles of the wave magnetic field $\bm{B}'$ and the velocity direction $\bm{e}'=\bm{e}$ are:
\bea
\tan\chi_B' &=& \frac{B_{w'}'}{B_{v'}'}=\frac{\left(\bm{k}'\times\bm{n}'\right)\cdot\bm{w}'}{\left(\bm{k}'\times\bm{n}'\right)\cdot\bm{v}'}, \\
\tan\chi_e' &=& \frac{e_{w'}}{e_{v'}}.
\eea
We go back to the $\mathcal{O}$ frame and find a basis, independent of that in the $\mathcal{O}'$ frame, in which the positional angle of the jet axis is 0:
\bea
\bm{v} &=& [-\cos\theta_{obs},0,\sin\theta_{obs}], \\
\bm{w} &=& [0,-1,0].
\eea
The positional angle of the velocity vector is
\be
\tan\chi_e=\frac{e_w}{e_v}=\frac{\sin\theta_e\sin\phi}{\sin\theta_e\cos\theta_{obs}\cos\phi-\cos\theta_e\sin\theta_{obs}}.
\ee
Finally, the EVPA is given by
\be
\chi_E=\chi_B'-\chi_e'+\chi_e-\frac{\pi}{2}.
\ee

For axisymmetric flows, the Stokes parameter $U_\nu$ averaged over the azimuthal angle $\phi$, for constant $z$ coordinate, vanishes. The averaged EVPA can be either $0$ (parallel polarization, $\left<Q_\nu\right>_\phi>0$) or $90^\circ$ (perpendicular polarization, $\left<Q_\nu\right>_\phi<0$). Thus we will use the quantity $\left<Q_\nu/I_\nu\right>_\phi$ as a polarization degree that contains information on the electric vector orientation.

\section{Conical shocks}
\label{sec:conical}

\citeauthor*{1990ApJ...350..536C} studied conical shocks as a mean of deflecting a relativistic upstream flow parallel to the jet axis. They kept the angle between the shock surface and the jet axis $\eta^{CC}\le\pi/2$. Thus, the matter crossed the shock from the outer side of the shock front (which is where the radial coordinate $r$ is larger than the shock surface radius for given $z$-coordinate) to the inner side. If $\eta^{CC}>\pi/2$, the situation is opposite: the upstream side is the inner side and the downstream side is the outer side. It has been discussed in a later paper \citep{2006MNRAS.367..851C}, that two conical shocks with $\eta^{CC}_1+\eta^{CC}_2=\pi$ and equal upstream velocities produces exactly the same Stokes parameters integrated over the azimuthal angle $\phi$. This type of symmetry is also evident in our procedure described in \S\ref{sec:oblique}. It can be easily shown, that transformation $\mathcal{T}=\{\eta\to-\eta, \theta_e\to-\theta_e, \phi\to\phi+\pi\}$ preserves $\mathcal{D}$, $\Pi$ and $\chi_E$, so it also preserves the Stokes parameters.

\begin{figure}
\includegraphics[width=\columnwidth,bb=60 70 290 180]{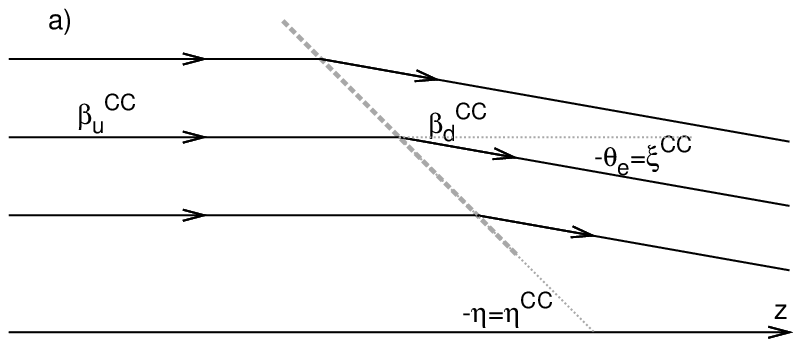}
\includegraphics[width=\columnwidth,bb=60 70 290 180]{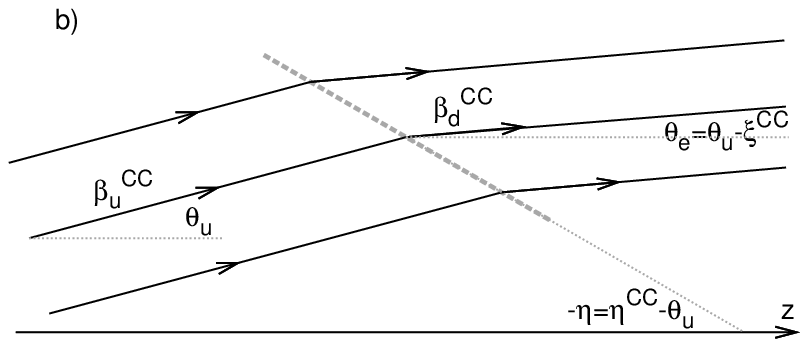}
\caption{Cross-sections through the conical shock: \emph{a)} with parallel upstream flow (like in \citeauthor*{1990ApJ...350..536C}), \emph{b)} with divergent upstream flow, where the shock surface (indicated by the \emph{thick gray dashed line}), the upstream and downstream velocity fields (\emph{solid black lines}) are all rotated by an angle $\theta_u$ with respect to the jet axis $z$. Thus, keeping the same value of upstream velocity $\beta_u^{CC}$, we obtain the same values for the downstream velocity $\beta_d^{CC}$ and the compression ratio $\kappa^{CC}$. $\eta^{CC}$ is the shock inclination to the upstream velocity field and $\xi^{CC}$ is the deflection angle between upstream and downstream velocity fields. The shock inclination $\eta$ is negative in both panels \emph{a)} and \emph{b)}, and downstream velocity field inclination $\theta_e$ is negative in panel \emph{a)}, while positive in panel \emph{b)}, assuming $\theta_u>\xi^{CC}$. As discussed in the text, the structure shown in panel \emph{a)} was obtained from a \citeauthor*{1990ApJ...350..536C} solution via transformation $\mathcal{T}$.}
\label{fig3.1}
\end{figure}

We calculate the polarization degree from conical shocks, when the upstream flow is not parallel, but diverges, making an angle $\theta_u$ with the jet axis. A model from \citeauthor*{1990ApJ...350..536C} with a shock inclination angle $\eta^{CC}$ and upstream velocity $\beta_u^{CC}$ is used for calculating the downstream velocity value $\beta_d^{CC}$, the downstream velocity deflection angle $\xi^{CC}$ and the shock compression ratio $\kappa^{CC}$ from their Eqs. (1), (5) and (2), respectively. For the procedure described in \S\ref{sec:oblique} we take: $\eta=\theta_u-\eta^{CC}$, $\theta_e=\theta_u-\xi^{CC}$, $\beta=\beta_d^{CC}$ and $\kappa=\kappa^{CC}$ (see Fig. \ref{fig3.1}). For $\theta_u=0$ this is equivalent to the \citeauthor*{1990ApJ...350..536C} solution with $\phi$ shifted by $\pi$.

\begin{figure}
\includegraphics[width=\columnwidth]{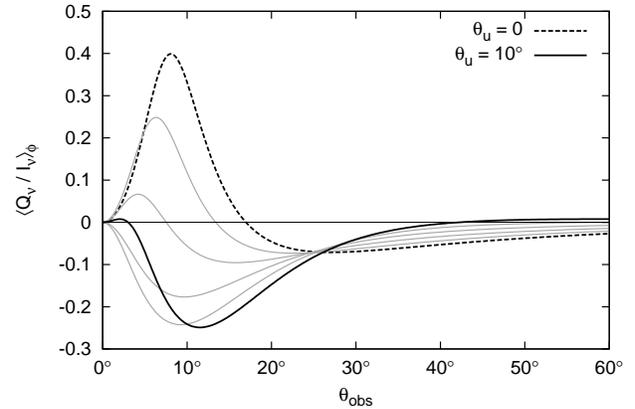}
\caption{Polarization degree from the conical shock averaged over the azimuthal angle $\phi$, as seen by different observers, calculated for models described in the text and in Fig. \ref{fig3.1}, based on the \citeauthor*{1990ApJ...350..536C} solution with the shock inclination angle $\eta^{CC}=10^\circ$ and the upstream Lorentz factor $\Gamma_u^{CC}=10$. Plotted are the results for $\theta_u=0$ (\emph{dashed line}), $\theta_u\in\{2^\circ,4^\circ,6^\circ,8^\circ\}$ (\emph{gray lines}) and $\theta_u=10^\circ$ (\emph{thick solid line}).}
\label{fig3.2}
\end{figure}

In Fig. \ref{fig3.2} we present the averaged polarization degrees, as the functions of observer inclination $\theta_{obs}$, for a series of models with $\eta^{CC}=10^\circ$, $\Gamma_u^{CC}=10$ and $\theta_u$ ranging from $0$ (\citeauthor*{1990ApJ...350..536C} solution) to $10^\circ$ ($\eta=0$, i.~e. a cylindrical shock surface). We find that for observers with $\theta_{obs}<25^\circ$, $\left<Q_\nu/I_\nu\right>_\phi$ strongly decreases with increasing $\theta_u$, changing sign from positive (parallel polarization) to negative (perpendicular polarization). For the cylindrical shock case a perpendicular polarization with degrees exceeding $20\%$ can be easily obtained. For observers with $\theta_{obs}>25^\circ$, $\left<Q_\nu/I_\nu\right>_\phi$ slightly increases with increasing $\theta_u$, nevertheless the polarization degrees are very low.

\begin{figure}
\includegraphics[width=\columnwidth]{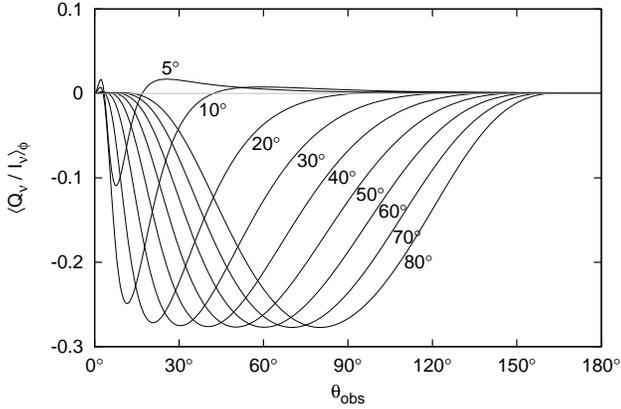}
\caption{Polarization degree from the cylindrical shock averaged over the azimuthal angle $\phi$, as seen by different observers. For each line the inclination angle of the upstream flow $\theta_u$ is indicated. All models were calculated for upstream Lorentz factor $\Gamma_u=10$.}
\label{fig3.3}
\end{figure}

We have found, that further increase of $\theta_u$ in the series does not result in deeper perpendicular polarization. Thus, we confine ourselves to study the cylindrical shock case with varying $\theta_u$. The results are shown in Fig. \ref{fig3.3}, which is analogous to Fig. 3b from \citeauthor*{1990ApJ...350..536C}. We find, that the polarization is dominantly perpendicular, with the polarization degree not exceeding $28\%$. The maximum polarization degree is very similar in all models with $\theta_u\ge 20^\circ$, it is approximately obtained for $\theta_{obs}\sim\theta_u$. Only for small $\theta_u$ some observers would see a parallel polarization of very low degrees.

\section{Reconfinement shocks}
\label{sec:reconf}

\begin{figure}
\includegraphics[width=\columnwidth]{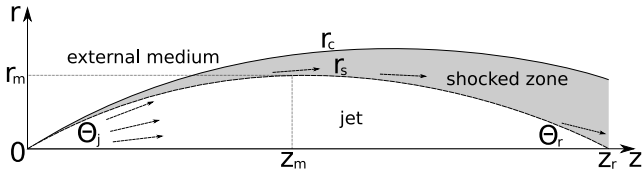}
\caption{Structure of the reconfinement shock for static external medium. The jet is launched at $z=0$ with half-opening angle $\Theta_j$. The jet matter (upstream) crosses the shock front described by $r_s(z)$ and gets slightly deflected in the downstream region. The shock reaches its maximum width $r_m$ at $z=z_m$ and ends in the recollimation point at $z=z_r\sim2z_m$ with the half-closing angle $\Theta_r\sim\Theta_j$. $r_c(z)$ denotes the contact discontinuity between the shocked jet matter and the external medium. This figure is taken from \citeauthor*{2009MNRAS.392.1205N}.}
\label{fig4.0.1}
\end{figure}

Following the idea that conical shocks can be approximated with a sequence of oblique shocks rotated around the jet axis \citep{1985ApJ...295..358L}, axisymmetric reconfinement shocks can be considered as a sequence of conical shocks with different inclination angle $\eta$. We use a semi-analytical model for the structure of reconfinement shocks from \citeauthor*{2009MNRAS.392.1205N} to provide downstream flow parameters for the polarization procedure described in \S\ref{sec:oblique}. Basic parameters of the reconfinement shock are presented in Fig. \ref{fig4.0.1}. Specifically, we take 'Model 2' described in that paper, which includes the transverse pressure gradient across the shocked matter zone, for the case of ultra-relativistic equation of state for the post-shock (downstream) matter ($\gamma_s=4/3$). It is characterized by: uniform external pressure $p_e$, bulk Lorentz factor of the jet $\Gamma_j$, half-opening angle of the jet $\Theta_j$ and total jet power $L_j$. As shown in \citeauthor*{2009MNRAS.392.1205N}, the parameters $p_e$ and $L_j$ have no influence on the shock shape, but they determine the size of the structure, which is represented by the reconfinement position $z=z_r$. We assume that the luminosity of the emitted synchrotron radiation is proportional to the kinetic energy flux dissipated at the given shock surface element.

\subsection{Polarization maps}
\label{subsec:reconf:resolv}

\begin{figure}
\includegraphics[width=\columnwidth,bb=64 98 306 201]{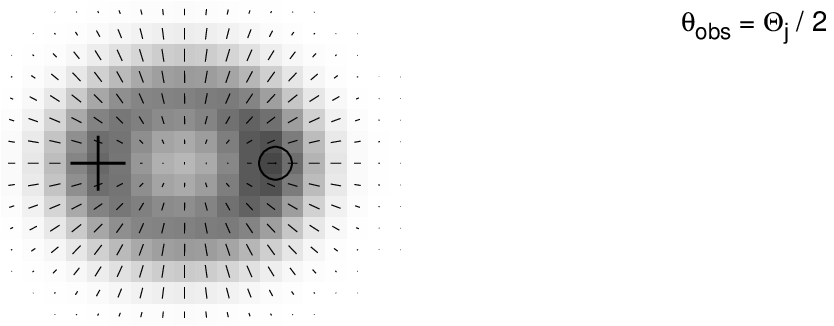}
\includegraphics[width=\columnwidth,bb=64 98 306 191]{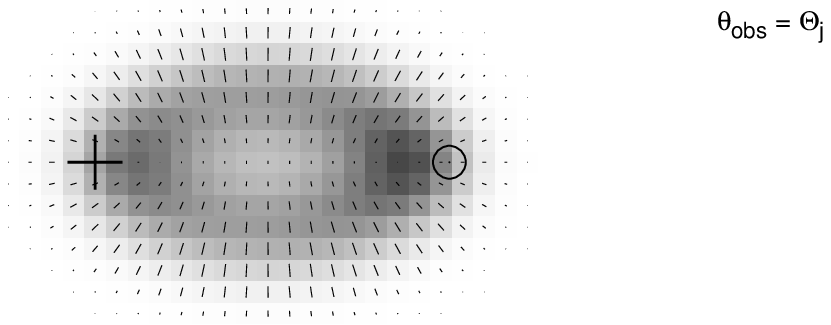}
\includegraphics[width=\columnwidth,bb=64 88 306 191]{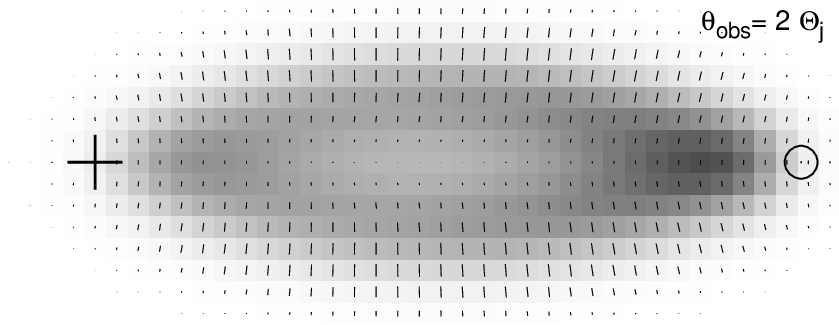}
\caption{Synthetic emission maps of the reconfinement shock with jet Lorentz factor $\Gamma_j=10$ and half-opening angle $\Theta_j=5^\circ$, as seen by different observers labelled with their inclination angle $\theta_{obs}$. The jet propagates horizontally to the right direction: from the source marked by \emph{the cross} to the recollimation point marked by \emph{the circle}. The linear scale and pixel resolution is the same in all cases. \emph{The gray shade} is proportional to the total observed intensity $I_\nu$, smoothed by a Gaussian distribution with standard deviation equal to the pixel diagonal size. The shading is normalized to the maximum pixel intensity calculated separately for each observer. Positional angle of the electric vector (EVPA) is indicated with \emph{bars} in the center of each pixel. The bar lenghts are proportional to the polarization degree. A bar of length equal to the pixel size would correspond to a $100\%$ polarized emission. The pixels of the lowest intensity were depolarized by adding a background of intensity equal to $3\%$ of the maximum pixel intensity.}
\label{fig4.1.1}
\end{figure}

In Fig. \ref{fig4.1.1} we show synthetic maps of synchrotron emission from spatially resolved reconfinement shock, a Model 2 solution with $\Gamma_j=10$ and $\Theta_j=5^\circ$, as seen by observers located inside ($\theta_{obs}=\Theta_j/2$), at ($\theta_{obs}=\Theta_j$) and outside ($\theta_{obs}=2\Theta_j$) the jet opening cone. We find that they would all see an edge-brightened jet. There are two regions of enhanced brightness. One is located close to the jet source (marked with crosses), although it is much weaker for $\theta_{obs}=2\Theta_j$. The other is close to the recollimation point (marked with circles). These regions are characterized with the strongest Doppler boost, as for some azimuthal angles the shock surface is approximately tangent to the line of sight.

The polarization degree maps are also edge-brightened, the polarization degree is especially small close to the jet axis. For small observer inclination ($\theta_{obs}=\Theta_j/2$), when the jet appears only slightly elongated, the polarization vectors point approximately radially outwards the midpoint between the jet source and the recollimation point. In the vicinity of the jet ends the polarization is parallel (to the jet axis), while between them it is perpendicular. For observer located at the jet opening cone ($\theta_{obs}=\Theta_j$) the polarization degree becomes lower, where polarization is parallel, while is high, where polarization is perpendicular. For large observer inclination ($\theta_{obs}=2\Theta_j$), the polarization is perpendicular everywhere in the structure.

\subsection{Longitudinal profiles}
\label{subsec:reconf:semiresolv}

When angular resolution is too low to resolve the transverse structure of the jet, one may still obtain a longitudal brightness and polarization profile. This is equivalent to integrating 2-dimensional maps like those shown in Fig. \ref{fig4.1.1} across the coordinate orthogonal to the jet projected axis. To compare such profiles obtained for differently oriented observers, we introduce a coordinate $\zeta$, which is related to the coordinates used in Fig. \ref{fig2.1} by
\be
\zeta=\frac{z-x\cot\theta_{obs}}{z_r}\,.
\ee
For points on the jet axis $x=0$, thus $\zeta=z/z_r$. For the recollimation point $\zeta=1$.

\begin{figure}
\includegraphics[width=\columnwidth]{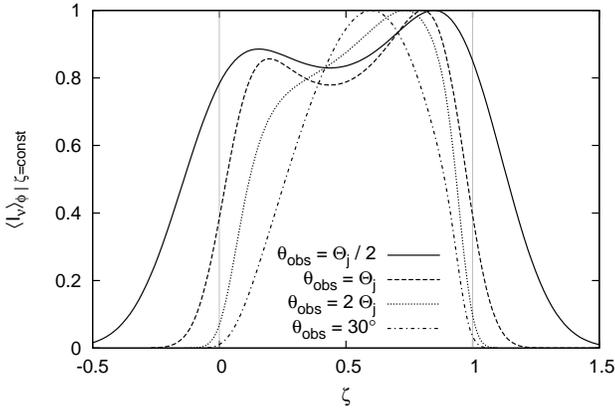}
\caption{Total observed emission intensity distribution along the jet direction, as seen by different observers labelled with their inclination angle $\theta_{obs}$. The profiles were obtained by integrating the emission maps similar to those shown in Fig. \ref{fig4.1.1} across the coordinate perpendicular to the projected jet axis. They are normalized separately to their maximum value. The $\zeta$-coordinate is defined in the text: $\zeta=0$ corresponds to the jet source and $\zeta=1$ to the recollimation point. All models were calculated for jet Lorentz factor $\Gamma_j=10$ and half-opening angle $\Theta_j=5^\circ$.}
\label{fig4.2.1}
\end{figure}

In Fig. \ref{fig4.2.1} we show the profiles of the total observed intensity for the same structure as in \S\ref{subsec:reconf:resolv}. Each profile has a clear maximum, which is close to the recollimation point for small $\theta_{obs}$ ($\zeta\sim0.85$ for $\theta_{obs}=\Theta_j/2$) and shifts towards the jet midpoint as $\theta_{obs}$ increases ($\zeta\sim0.6$ for $\theta_{obs}=30^\circ$). A secondary maximum appears close to the jet source ($\zeta\sim0.2$) for $\theta_{obs}$ smaller (and apparently also for slightly larger) than $\Theta_j$. Thus the profile changes from double-peaked to approximately parabolic, as $\theta_{obs}$ increases. This is consistent with total intensity maps from Fig. \ref{fig4.1.1}. This profile morphology change reflects the importance of relativistic boosting of the radiation. As was shown in Fig. 9 of \citeauthor*{2009MNRAS.392.1205N}, most of the kinetic energy dissipation takes place for $\zeta\sim0.5$. The luminosity profiles for large $\theta_{obs}$ are consistent with this, but the profiles for $\theta_{obs}\le\Theta_j$ show effects of strong Doppler boosting from the jet portions characterized with $|\eta|\sim\theta_{obs}$, thus producing peaks that do not correspond to the position of maximum kinetic energy dissipation.

\begin{figure}
\includegraphics[width=\columnwidth]{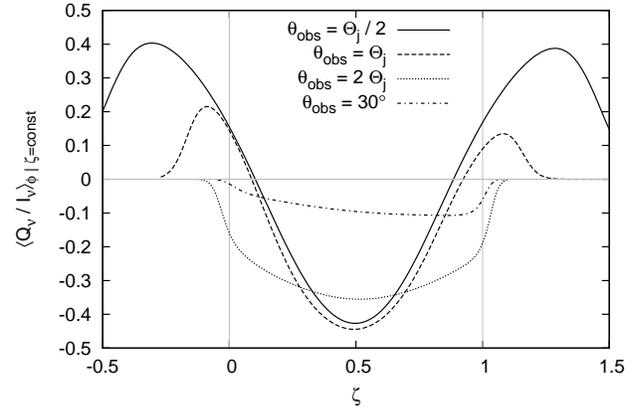}
\caption{Polarization degree profiles along the jet direction, as seen by different observers labelled with their inclination angle $\theta_{obs}$. They were obtained from the Stokes parameters $Q_\nu$ and $I_\nu$ integrated in the way described in Fig. \ref{fig4.2.1}. All models were calculated for jet Lorentz factor $\Gamma_j=10$ and half-opening angle $\Theta_j=5^\circ$.}
\label{fig4.2.2}
\end{figure}

In Fig. \ref{fig4.2.2} we show the polarization degree profiles for the same structure. Observers with $\theta_{obs}\le\Theta_j$ would see parallel polarization from regions with $\zeta<0.1$ and $\zeta>0.9$ and perpendicular otherwise. The maximum parallel polarization degree is obtained outside both projected ends of the structure ($\zeta<0$ and $\zeta>1$) and is higher for $\theta_{obs}=\Theta_j/2$ ($\sim40\%$) than for $\theta_{obs}=\Theta_j$ ($\sim20\%$). The maximum perpendicular polarization degree ($\sim40\%$) corresponds to $\zeta\in[0.4;0.6]$.

For $\theta_{obs}>\Theta_j$ the observers see perpendicular polarization for all $\zeta$. The polarization is only significant between the ends of the structure, namely for $0<\zeta<1$. The maximum polarization degree is decreasing with increasing $\theta_{obs}$, changing from $\sim-35\%$ for $\theta_{obs}=2\Theta_j$ to $\sim-10\%$ for $\theta_{obs}=30^\circ$. The polarization degree profile changes from approximately symmetric with respect to $\zeta=0.5$ for $\theta_{obs}=2\Theta_j$ to that with the polarization degree increasing linearly with $\zeta$ for $\theta_{obs}=30^\circ$.

Note that all of the observers see perpendicular polarization from the region with $\zeta\in[0.1;0.9]$. In this region they see dominantly the portions of the shock with small inclinations $|\eta|\lesssim\Theta_j/2$. Thus, the perpendicular polarization seen there is consistent with the cylindrical shock solutions discussed in \S\ref{sec:conical}. As one can see in Fig. \ref{fig4.2.1}, this is also the region, from which the bulk of the total emission originates. Thus, we expect that this region dominates the radiative output from the reconfinement shock and so the averaged polarization from the whole structure is perpendicular for most observers.

\subsection{Spatially averaged polarization}
\label{subsec:reconf:unresolv}

\begin{figure}
\includegraphics[width=\columnwidth]{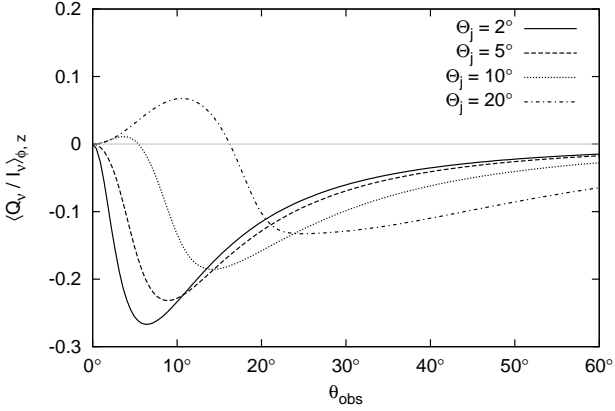}
\caption{Spatially averaged polarization degrees from the reconfinement shock, as seen by different observers labelled with their inclination angle $\theta_{obs}$. Models were calculated for jet Lorentz factor $\Gamma_j=10$ and different jet half-opening angles $\Theta_j$.}
\label{fig4.3.1}
\end{figure}

In Fig. \ref{fig4.3.1} we show the polarization degree averaged over the whole spatial extent of the reconfinement shock for four models with $\Gamma_j=10$ and different half-opening angles $\Theta_j$. The polarization is dominantly perpendicular, with maximum degree observed for $\theta_{obs}\gtrsim\Theta_j$. The maximum polarization degree value is decreasing with $\Theta_j$, ranging from $\sim27\%$ for $\Theta_j=2^\circ$ to $\sim13\%$ for $\Theta_j=20^\circ$. For $\Theta_j>\Gamma_j^{-1}$ observers closely aligned with the jet axis will see some parallel polarization, with maximum polarization degree $\sim7\%$ for $\Theta_j=20^\circ$. For observers with large $\theta_{obs}$ the polarization degree increases with $\Theta_j$.

\begin{figure}
\includegraphics[width=\columnwidth]{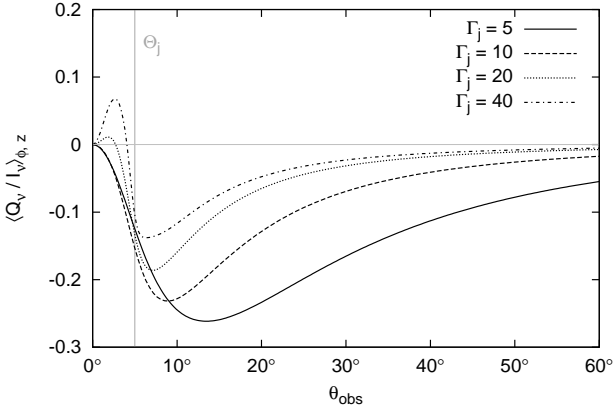}
\caption{Spatially averaged polarization degrees from the reconfinement shock, as seen by different observers labelled with their inclination angle $\theta_{obs}$. Models were calculated for jet half-opening angle $\Theta_j=5^\circ$ (marked with the \emph{gray vertical line}) and different jet bulk Lorentz factors $\Gamma_j$.}
\label{fig4.3.2}
\end{figure}

In Fig. \ref{fig4.3.2} we show the spatially averaged polarization degree from the reconfinement shock for four models with $\Theta_j=5^\circ$ and different bulk Lorentz factors $\Gamma_j$. Again, the polarization is mostly perpendicular, with the exception of closely aligned observers, when $\Gamma_j>\Theta_j^{-1}$ (with maximum parallel polarization degree $\sim7\%$ for $\Gamma_j=40$). The position of maximum polarization degree approaches $\Theta_j$ with increasing $\Gamma_j$ and the maximum polarization degree value decreases, ranging from $\sim26\%$ for $\Gamma_j=5$ to $\sim14\%$ for $\Gamma_j=40$. For observers with large $\theta_{obs}$ the polarization degree significantly decreases with $\Gamma_j$.

\section{Discussion \& Conclusions}
\label{sec:discussion}

We have shown that the degree of perpendicular polarization from relativistic conical shocks with chaotic magnetic fields can be large, if one considers divergent upstream velocity field. In the extreme case, when the shock surface is parallel to the jet axis (a cylindrical jet) the degrees of perpendicular polarization may reach $28\%$. Thus, the conical shocks can account for the observed polarization bimodality of parsec-scale jets on mm wavelengths \citep{1998MNRAS.297..667N, 2007AJ....134..799J}, without introducing parallel magnetic fields, as proposed in \cite{2006MNRAS.367..851C}. It is important to notice, that one can only consider 'local' solutions of this kind, i.~e. axisymmetric sections that are far enough from the axis, to avoid problems on the $r=0$ boundary.

Our results on polarization from the conical shocks have important implications for the reconfinement shocks, that are modelled in terms of an interaction between the spherically divergent relativistic jet (upstream) flow and the static external medium. Every particular section of approximately paraboloidal shock surface can be obtained from a \citeauthor*{1990ApJ...350..536C} solution via simple rotation described in Fig. \ref{fig3.1}. The main difference between our and their models of the shock configuration is that we assume a cold upstream matter, making possible the solutions for arbitrarily small inclinations of the shock front, with respect to the upstream velocity field.

We have found, that the emission from axisymmetric reconfinement shocks with chaotic magnetic fields is clearly dominated by perpendicular polarization. Both total intensity and polarization degrees are higher on the edges of the structure and the wave electric vectors are perpendicular to the outline of the shock surface. The polarization maps are axially symmetric, in accordance with the absence of a large-scale helical magnetic field component \citep{2005MNRAS.360..869L}. The total intensity maps are strongly affected by Doppler boosting for observers located closely to the jet opening cone. One can see two 'knots' for $\theta_{obs}\in\{\Theta_j/2,\Theta_j\}$ close to both of the shock ending points, but only one knot for $\theta_{obs}=2\Theta_j$ close to the recollimation point. On the longitudinal profiles (Fig. \ref{fig4.2.1}) we observe this knot to shift towards the shock midpoint and loose its brightness contrast. The polarization of the knots can be estimated, by summing the Stokes parameters from the 3x3 pixel groups taken from Fig. \ref{fig4.1.1} and centered on the brightest pixel of the knot. It is parallel with degrees about $30\%$ for both knots seen for $\theta_{obs}=\Theta_j/2$. The knots seen for $\theta_{obs}=\Theta_j$ are effectively depolarized, the one close to the source of the jet having $4\%$ net parallel polarization, while the other one showing $1\%$ net perpendicular polarization. Finally, the knot seen for $\theta_{obs}=2\Theta_j$ shows perpendicular polarization with degree of $23\%$. Thus we predict that the polarization of the knots associated with the reconfinement shocks may be both parallel or perpendicular, depending on the observer position with respect to the jet opening cone.

The longitudinal profiles of the polarization degree (Fig. \ref{fig4.2.2}) show that observers closely aligned with the jet axis may see polarization degrees as high as $40\%$, both parallel and perpendicular. This is higher than the values obtained for conical shocks. It can be explained as a projection effect, since points of equal $\zeta$ coordinate form a plane oblique to the jet axis. Integrating emission from such planes includes shock surface portions with different inclination to the jet axis. If we instead integrated emission from the points of equal $z$ coordinate, we would obtain polarization degrees consistent with the solutions for conical shocks.

On the longitudinal profiles for closely aligned observers, the transition between the parallel and perpendicular polarization is approximately coincident with the total intensity profile maxima. Note, that for $\theta_{obs}=\Theta_j/2$ the total intensity profile maxima are slightly shifted towards the jet midpoint from the positions of the highly parallel-polarized knots.

The spatially averaged polarization degrees also prove the dominance of perpendicular polarization. The models with $\Gamma_j\Theta_j<1$ produce perpendicular polarization for all observers, while the models with $\Gamma_j\Theta_j>1$ can produce parallel polarization only for the observers located inside the jet opening cone. This results from very strong Doppler boosting from the shock surface regions tangential to the line of sight. The transition between polarization degree maximum and minimum always takes place around $\theta_{obs}=\Theta_j$ and its sharpness $-d(\left<Q_\nu/I_\nu\right>_{\phi,z})/d\theta_{obs}$ increases with $\Gamma_j$. Thus, extremely relativistic reconfinement shocks could produce the polarization angle swings with moderate change of jet orientation.

\section*{Acknowledgments}

The author thanks Marek Sikora for invaluable discussions, advices and comments on the early versions of the manuscript. An anonymous referee provided some very helpful remarks. This work was partially supported by the Polish MNiSW grant N N203 301635 and the Polish Astroparticle Network 621/E-78/BWSN-0068/2008.

\appendix

\section{Lorentz invariance of the polarization angle}
\label{app1}

It is well known that the polarization angle of linearly polarized radiation is invariant in Lorentz boost, if this angle is measured from the plane containing both the wave propagation direction and the boost velocity vector \citep{1972Natur.240..161C}. Straightforward formulae for the value of polarization angle can be derived, when one chooses to set up his coordinate system aligned with this preferred plane \citep{1979ApJ...232...34B, 1982ApJ...260..855B}. However it is sometimes more convenient to use a fixed coordinate system, in which the wave propagation direction and the boost direction are completely arbitrary. One has also a freedom of choosing a basis in the plane of the sky, in which polarization angle is associated with the azimuthal angle of the wave electric vector. The formulae for the synchrotron polarization angle for arbitrary basis and magnetic field direction have been given by \cite{2003ApJ...597..998L} (see their Appendix C). If the plane-of-the-sky basis is chosen to be somehow aligned with the fixed coordinate system, the polarization angle has to be transformed in non-trivial way. But in this transformation it this also the basis that rotates, and so the projection of the velocity vector onto the basis is different. Below we show in most general approach, that the angle between the polarization vector and the projection of the transformation velocity vector onto the plane of the sky is Lorentz invariant.

Consider a frame $\mathcal{O}$, in which a linearly polarized wave propagates towards an observer pointed by unit vector $\bm{k}$. Frame $\mathcal{O}'$ is moving with the velocity $\bm\beta=\beta\bm{n}$, where $|\bm{n}|=1$. The polarization vector positional angle $\chi$ of the wave is measured in the plane of the sky, begining from $\bm{n}$ projected onto the plane of the sky. We construct an orthogonal basis ($\bm{v}$, $\bm{w}$) in the plane of the sky, that fulfills the conditions $\bm{n}\cdot\bm{v}>0$ and $\bm{n}\cdot\bm{w}=0$:
\bea
\bm{v} &=& \frac{\bm{n}-(\bm{n}\cdot\bm{k})\bm{k}}{\sqrt{1-(\bm{n}\cdot\bm{k})^2}},\\
\bm{w} &=& \bm{k}\times\bm{v}=\frac{\bm{k}\times\bm{n}}{\sqrt{1-(\bm{n}\cdot\bm{k})^2}}.
\eea
The electric vector of the wave $\bm{E}=E\bm{e}$ has its direction determined by the polarization angle:
\be
\bm{e}=\cos\chi\,\bm{v}+\sin\chi\,\bm{w}.
\ee
The magnetic vector of the wave is $\bm{B}=B\bm{b}$, where $B=E$ and $\bm{b}=\bm{k}\times\bm{e}$. The Lorentz transformation of electric field is given by \citep{1979rpa..book.....R}
\bea
E_\parallel' &=& E_\parallel,\\
\bm{E}_\perp' &=& \Gamma\left(\bm{E}_\perp+\bm\beta\times\bm{B}\right).
\eea
Hence the electric vector in $\mathcal{O'}$ is
\be
\bm{E}'=\Gamma\bm{E}-(\Gamma-1)(\bm{n}\cdot\bm{E})\bm{n}+\Gamma\beta\,\bm{n}\times\bm{B}.
\ee
One can show, that
\be
E'=\Gamma\left(1-\beta\,\bm{n}\cdot\bm{k}\right)E=\frac{E}{\mathcal{D}}.
\ee
The field direction can be written as
\be
\label{vecep}
\bm{e}'=\frac{\cos\chi\,\bm{l}+\sin\chi\,\bm{k}\times\bm{n}}{\sqrt{1-(\bm{n}\cdot\bm{k})^2}}\,,
\ee
where
\be
\bm{l}=\left[1-\mathcal{D}(\Gamma-1)\left(1-(\bm{n}\cdot\bm{k})^2\right)\right]\bm{n}-\left[\frac{\bm{n}\cdot\bm{k}-\beta}{1-\beta\,\bm{n}\cdot\bm{k}}\right]\bm{k}.
\ee
Note, that $\bm{l}\cdot\left(\bm{k}\times\bm{n}\right)=0$ and $\left|\bm{l}\right|=\sqrt{1-(\bm{n}\cdot\bm{k})^2}$, so in Eq. (\ref{vecep}) we have effectively decomposed the transformed electric field direction in a new orthogonal basis. Thus, we conclude that
\be
\tan\chi'=\tan\chi\,,
\ee
i.~e. the polarization angles are the same in $\mathcal{O}$ and $\mathcal{O}'$.

\end{document}